\documentclass[a4paper]{article}

\usepackage{INTERSPEECH2020}
\usepackage{algorithm}
\usepackage{algorithmic}
\usepackage{multirow}
\usepackage{booktabs}
\usepackage{epsfig}
\usepackage{subfigure}
\usepackage{subfig}

\title{Evolutionary Algorithm Enhanced Neural Architecture Search for Text-Independent Speaker Verification}

\name{Xiaoyang Qu, Jianzong Wang$^*$, Jing Xiao\thanks{*Corresponding author: Jianzong Wang, jzwang@188.com}}
\address{
  Ping An Technology (Shenzhen) Co., Ltd.}
\email{\{quxiaoyang343,wangjianzong347,xiaojing661\}@pingan.com.cn}

\begin{document}

\maketitle
\begin{abstract}
State-of-the-art speaker verification models are based on deep learning techniques, which heavily depend on the hand-designed neural architectures from experts or engineers. We borrow the idea of \textit{neural architecture search(NAS)} for the \textit{text-independent speaker verification task}. As NAS can learn deep network structures automatically, we introduce the NAS conception into the well-known x-vector network. Furthermore, this paper proposes an evolutionary algorithm enhanced neural architecture search method called Auto-Vector to automatically discover promising networks for the speaker verification task. The experimental results demonstrate our NAS-based model outperforms state-of-the-art speaker verification models.
\end{abstract}
\noindent\textbf{Index Terms}: speaker verification, deep learning, neural network, neural architecture search.

\section{Introduction}
\label{sec:intro}
Speaker verification is the process of verifying whether an utterance belongs to the same speaker, based on enrolled speaker information. It can be categorized as \textit{text-dependent speaker verification} (TD-SV) and \textit{text-independent speaker verification} (TI-SV).  Relatively, TI-SV is more convenient for practical applications, as it poses no constraints, e.g., duration or lexical content, on utterances to verify. However, it is also more difficult to achieve a good performance, due to many potential variabilities of the utterances.  In this work, we focus on TI-SV.

In the early years, the i-vector \cite{ivector} based models with PLDA\cite{PLDA} backend dominated the development of the speaker verification application. In recent years, the deep neural networks(DNN) trained as acoustic models for automatic speech recognition (ASR) are integrated into the i-vector system\cite{ASRDNN1,ASRDNN2,ASRDNN3}. Although the ASR DNN can enhance phonetic modeling in the i-vector UBM, it adds a high computational cost to the i-vector system.  In the latest years, DL-based techniques can be used as utterance-level speaker feature extractor\cite{Dvector-smallTDSV,DvectorTDSV,DoubleJVector,Tandem-TDSV}, and enable an end-to-end pipeline to discriminate between speakers\cite{E2E,GE2E,DeepSpeaker}.


However, these architectures are hand-designed by experts or experienced engineers. It is highly demanding on their knowledge and experiences. As a result, neural architecture search\cite{RL_NAS_START1}\cite{RL_NAS_START2} techniques are becoming an increasingly popular topic in both academia and industry, because of its great potential to automatically find more effective architectures to outperform hand-crafted ones. The early works on NAS are based on reinforcement learning or evolutionary algorithm, such as \cite{RL_NAS_START1,NEAT,Ameoba,EvolvingDNN,MemeticEvolution} . But these approaches are expensive in time. To reduce search time costs,  researchers proposed a wide range of optimization paradigms\cite{PNAS,DeepArchitect,ENAS,morphing}, where hyper-network\cite{SMASH, One-shot,GNNNAS,SinglePath} is a typical representative. 



In this work, we bring the idea of hyper-network based neural architecture search into text-independent speaker verification. We managed to improve its search efficiency by use of a memetic evolutionary algorithm. Our work has several contributions as follows. (1) As NAS can learn deep network structures automatically, we introduce the NAS conception into the x-vector network. (2) To learn more promising structures for speaker verification, we build a large-scale hyper-network with repetitive architecture motifs. (3) To discover more promising candidate networks, we use a memetic evolutionary algorithm. (4) The experiment results demonstrate that our NAS-based x-vector and Auto-Vector outperform state-of-the-art speaker verification methods in two datasets. 


\begin{figure*}[!htb]
  \centering
    \includegraphics[width=0.9\textwidth]{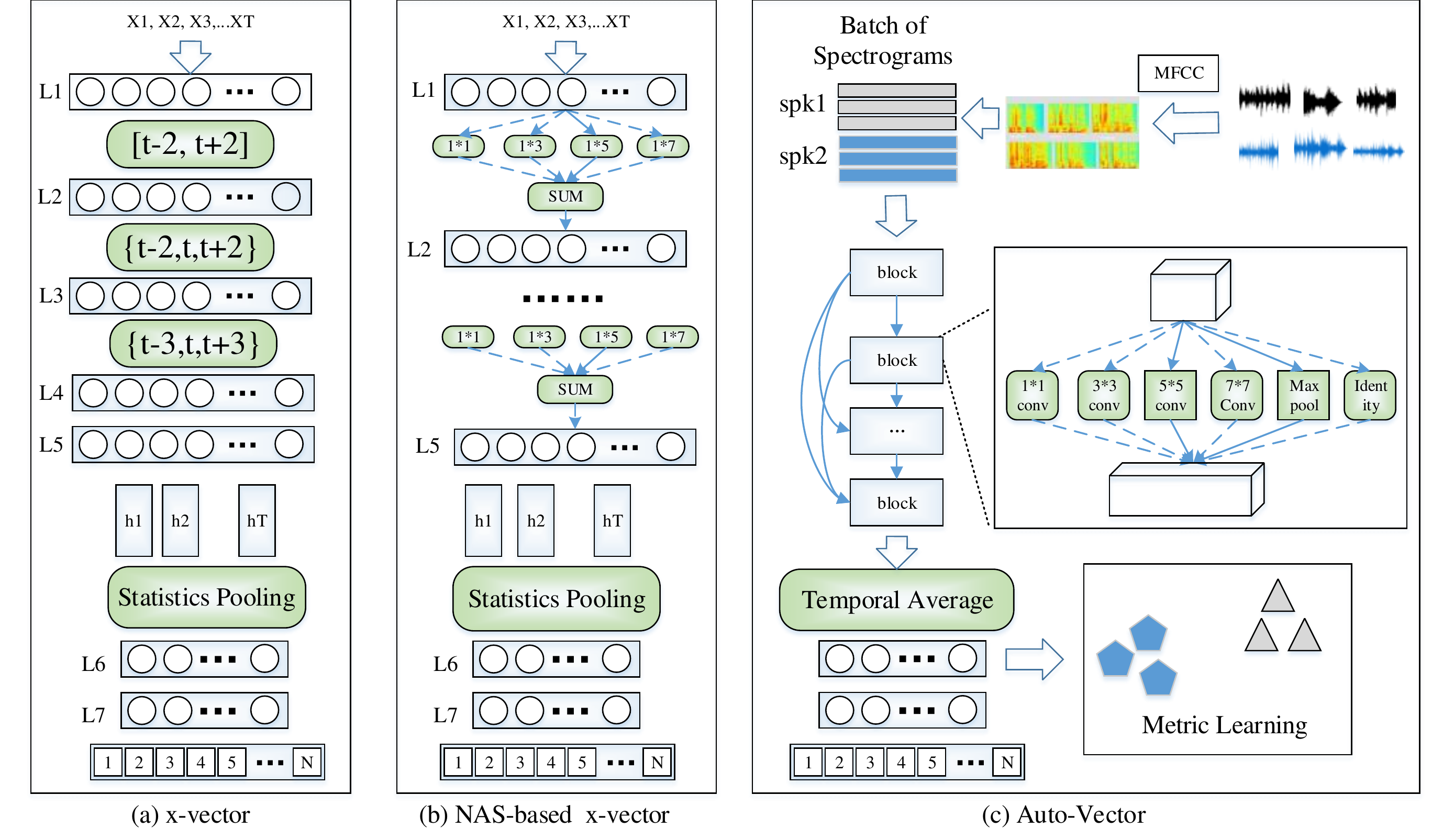}
    \caption{(a)The embedding DNN architecture of x-vector. (b)Our NAS-based x-vecotr. (c)Our Auto-Vector for speaker verification. } 
    \label{fig:mainidea}
\end{figure*}

\section{Proposed Methods}

\subsection{NAS-based x-vector}

First, let us review the well-known x-vector network as shown in Figure\ref{fig:mainidea}(a). Suppose the input utterance contains T frames. The first five layers $L_1$ to $L_5$  are frame-level information hidden layers. These layers are connected with a time-delay architecture with temporal context windows. The context window over the first layer is set as a range from $T-2$ to $T+2$. The second and third layers splice the output of the previous layer at time steps \{T-2, T, T+2\} and \{T-3, T, T+3\}, respectively. The statistical pooling layer builds utterance-level feature by calculating mean and standard deviation over frame-level features. Note, the seventh hidden layer $L_7$ and the final softmax output layer are used for training and discarded in the evaluation process. The sixth layer $L_6$ is used as the embedding of x-vector. 

As NAS can learn deep network structures automatically, we introduce the NAS conception into the x-vector network, as shown in Figure\ref{fig:mainidea}(b). For conventional x-vectors, the context windows between frame-level hidden layers are set by experts. Here, we let the number and the size of the context window to be decided by an automatic-decided method. This method is developed from hyper-network-based NAS, which stands out among these efficient NAS approaches because they can significantly reduce the tedious training process by sharing its parameters with all candidate networks. The key point is to specify the search space of hyper-network, which contains all possible candidate networks. As shown in Figure \ref{fig:mainidea}(b), we incorporate various choice temporal context windows for the first five layers.  Then, we use a memetic evolution search policy shown in Section \ref{sec:searchpolicy} to find the optimal candidate network with combination choices for temporal context windows. The statical pooling, the sixth, and seventh layers are the same as conventional x-vectors. However, the small search space limits the potentiality of NAS-based x-vector. To enable an ample search space, we designed an Auto-Vector for speaker verification, as shown in Section\ref{sec:autovector}.


\subsection{Auto-Vector}\label{sec:autovector}

To automatically learn more promising nerual architectures for text-independent speaker verification, we build a large-scale hyper-network with repetitive architecture motifs. As shown in Figure\ref{fig:mainidea}(c), the framework includes three parts: input features, architecture, and loss. 


\textbf{Input Features}. MFCCs( Mel-frequency cepstral coefficients) is used to extract the frame-level acoustic feature vectors from raw waveform signals. Then the frames are converted into input acoustic features of 40-dimensional MFCCs with a frame-length of 25ms. This gives spectrograms of size 40*300 for 3 seconds of speech.

\textbf{Architecture}. As shown in Figure \ref{fig:mainidea}(c), we build a hyper-network containing the entire search space of architectures. The architecture of hyper-network is stacked with identical structure but different weights. Assume there are $N_B$ choice blocks, and every block has $N_{op}$ choice operations. Each choice block applies either one or two different operations out of $N_{op}$ possible options. Thus, there are, therefore $N_{op}+\frac{N_{op}*(N_{op}-1)}{2}$ possible combinations of operations that we can apply in each block. In our experiment, the $N_{op}$ is set as 6, so we have six possible operations:  a max-pooling layer, an identity operation, and convolution layers of size 1x1, 3x3, 5x5 and 7x7. The average temporal pooling is implemented by applying a 2D adaptive average pooling several input planes.  As the size of the search space grows exponentially with the number of choice blocks  $N_B$, this large-scale search space can enable more possibility of promising networks.

\textbf{Loss.} In addition to softmax pre-training, we also use distance-based loss function, such as triplet loss or generalized end-to-end loss. Among various distance-based loss functions \cite{DeepSpeaker, GE2E, E2E}, the generalized end-to-end loss function \cite{GE2E} perform best, because it not only learns to rank but also emphasizes the hard examples. The details are shown in Section \ref{sec:backend}.


\subsection{Evolutionary Algorithm Enhanced NAS}

\subsubsection{The Overall Training Procedure}
The training procedure of NAS-based x-vector and Auto-Vector consist of four steps: (1) Design a search space. (2) Train the hyper-network. (3) Search the optional sub-networks with their parameters inherited from the hyper-network. (4) Retrain the best-accuracy candidate sub-network as a standalone model. The \textit{first step} have been shown in Section\ref{sec:autovector}. The \textit{second step} is to train hyper-network. The training goal is formulated as
\begin{equation}
     \theta^*(H)=\underset{\theta}{argmin}\: \mathcal{L}_{train}(H,\theta)
\end{equation}
here $\theta^*$ means the weights of the hyper-network $H$ and $\mathcal{L}_{train}(\cdot)$ denotes a loss function on training dataset.

The \textit{third step} is to search for high-quality sub-networks, with their parameters inherited from the hyper-network.The high-quality sub-network search task is a black-box optimization, which aims to find an approximate maximizer of an objective function f(x) using a given budget of N sub-network evaluations. In can be formulated as 
\begin{equation}
\tilde{a}^*\approx a^* = \underset{a_n}{argmax}f(a_n \sim H) 
\end{equation}
note the sub-network $a_n$ is sampling from the search space of hyper-network $H$. For this optimization goal, we develop a memetic algorithm based evolutionary policy, which is illustrated in Section\ref{sec:searchpolicy}, 

The \textit{last step} is to retrain the obtained optimal sub-network for the best performance. The model parameters learned by minimizing the accumulative loss shown in the equation of 
\begin{equation}
    \phi ^*(\tilde{a}^*)=\underset{\phi}{argmin}\: \mathcal{L}_{train}(\tilde{a}^*,\phi)
\end{equation}
here $\phi^*$ means the weights of the best-accuracy candidate sub-model $\tilde{a}^*$. And,  $\mathcal{L}_{train}(\cdot)$ denotes a loss function on training dataset. The details of loss function is shown in Section\ref{sec:backend}.

\subsubsection{Memetic Evolutionary Search Policy}\label{sec:searchpolicy}

Our search policy is based on the memetic algorithm. The memetic algorithm is an augmentation of the genetic algorithm. In other words, the memetic algorithm consists of the genetic algorithm and one or more local search components. The memetic algorithm integrates the local search method into the genetic algorithm to reduce the likelihood of premature convergence. Thereby, the promising child individuals are generated by recombination from and adaptation from outstanding individuals. 

The search process is shown in Algorithm \ref{alg:memetic}. The inputs include an empty population set $\Omega$ with size $S$, the generation number $G$, and the well-trained hyper-network $H$. The key operations of the search process are shown as follows. (1) For \textit{mutation operations}, the selected candidate choose one or two different operations in its every choice block with probability 0.1 to produce a new candidate. Because cross-over operations will result in local operation, we only make mutation operations.  (2) The \textit{local search} employs a hill-climbing algorithm to discover high-quality sub-networks by greedily moving in the direction of better-performing sub-networks. (3) The \textit{compete operation} uses an acceptance criterion to pick the better one. (4) The \textit{fitness evaluation} is calculating as $f_i=1-\delta_i$, where $\delta_k$ means the equal error rate of $i$ individual model. (5) The \textit{selection operation} is based on a tournament selection policy. In the tournament selection policy, a candidate set is randomly selected from the overall population set. Then,  the best-fitness individual is chosen from the candidate set rather than the overall population set. This policy can avoid zooming in on good models too early and enable more search space to be explored.

\begin{algorithm}  
\caption{Memetic evolutionary algorithm}  
\label{alg:memetic}  
\begin{algorithmic}[1] 
\STATE \textbf{Input:} The population set $\Omega$ with size $S$ 
\STATE \textbf{Input:} The generation number $G$
\STATE \textbf{Input:} The well-trained hyper-network $H$
\STATE \textbf{Output:} A best-accuracy sub-network $a$
\STATE Initialize population set $\Omega$
\FOR{ $i \leftarrow  1...S$}
\STATE $a_i \leftarrow$ uniform-sampling($H$)
\STATE $f_i \leftarrow$ fitness evaluation($a_i$)
\STATE  $ \Omega =  \Omega + \: \{a_i,f_i\}$
\ENDFOR
\FOR{ $j \leftarrow  1...G$}
\STATE  $\ddot{a_j} \leftarrow$ mutation($a_j$)
\STATE  $\bar{a_j} \leftarrow$ local-search($\ddot{a_j}$)
\STATE  $a_j \leftarrow$ compete($\ddot{a_j},\bar{a_j}$)
\STATE  $f_j \leftarrow$ fitness-evaluation($a_j$)
\STATE  $ \Omega =  \Omega + \: \{a_j,f_j\}$  \# add the promising candidate
\STATE  $ \Omega =  \Omega -$ worst($\Omega$)  \# remove the worst candidate.
\STATE  $a_{j+1} \leftarrow$ tournament selection($\Omega$)
\ENDFOR
\end{algorithmic}  
\end{algorithm}

\subsection{Backend}\label{sec:backend}
The objective of typical cross-entropy loss is to learn to predict directly a label given an input. Metric learning aims to predict relative distance between inputs. In addition to softmax pre-training, we also use distance-based loss function. Assuming N speakers with each M utterance. The loss function $\mathcal{L}(\cdot)$ is shown as follows. 
\begin{equation}
    \mathcal{L}(e_a,e_p,e_n)=1-\sigma (d(e_a,e_p))+\underset{\substack{1\leq k\leq N\\k\neq j}}{max}\sigma (d(e_a,e_n^k))
\end{equation}
where $d(e_a,e_p)$ means the scaled similarity score between the anchor embedding $e_a$ and the positive embedding $e_p$. Here, $e_a$ and $e_p$ belongs to the same speaker. The negative embedding $e_n^k$ is the centroid embedding of the $k_{th}$ speakers, which should be evaluated as $e_n^k = \frac{1}{M}\sum_{m=1}^{M}e_{a}^k(m)$, using M utterances for the $k_{th}$ speaker. $\sigma(\cdot)$ means the sigmoid function. Here, the scaled similarity score function $d(e_a,e_p)$ is defined as
\begin{equation}
    d(e_a,e_p)=w\cdot cos(e_a,e_p)+b
\end{equation}
here, $w$ and $b$ are learnable parameters. $cos(\cdot)$ means the cosine similarity function. 

\section{EXPERIMENT}

\subsection{Dataset Collection and Pre-Processing}
We use two datasets for Evaluation. \textbf{Dataset1} includes 300 speakers with 4527 utterances in total. The duration of which mostly range from 3 to 7 seconds. We split the overall dataset into a training dataset of 270 speakers and a test dataset of 30 speakers. 10 utterances are randomly chosen as enrollment utterances for each speaker, and another 10 randomly chosen utterances are used as evaluation samples. 

\textbf{Dataset2} includes 4000 speakers with 23573 utterances and more than 12,600 hours of speech. This dataset is split into two parts: a training dataset of 3960 speakers and an evaluation dataset of 285 speakers. The evaluation partition consists of 285 speakers that do not overlap with the 3960 speakers for training datasets.  

The raw waveform audios with a 16KHz sampling rate are converted into frames using a hamming window of width 25 ms and step 10ms. MFCCs( Mel-frequency cepstral coefficients) is used to extract the frame-level acoustic feature vectors from raw waveform signals. Then the frames are converted into input acoustic features of 40-dimensional MFCCs with a frame-length of 25ms that are mean-normalized over a sliding window of up to 3 seconds. This gives spectrograms of size 40*300 for 3 seconds of speech. An energy-based VAD is employed to filter out non-speech frames from the utterances. There are N speakers each with M utterances.

\subsection{Overall Result}

\begin{table}
\centering
\caption{Equal Error Rate Comparison(The Lower, The Better)}
\label{table:overall-eer}
\begin{tabular}{cccc}
\toprule   
& \begin{tabular}[c]{@{}c@{}}EER\\ (Dataset1)\end{tabular} &\begin{tabular}[c]{@{}c@{}}EER\\ (Dataset2)\end{tabular} & Size \\
\midrule   
LSTM-GE2E\cite{GE2E} & 6.2\% & 8.3\% & 4.6M\\
x-vector\cite{Xvector} & 4.6\% & 6.5\%  & 6.14M \\
NAS-based x-vector & 4.3\% & 5.6\% & 6.32M \\
Auto-Vector & \textbf{1.8\%} & \textbf{3.6\%} & 5.17 M \\
\bottomrule   
\end{tabular}
\end{table}

Table \ref{table:overall-eer} shows the EER comparison of four models on Dataset1 and Dataset2. Our Auto-Vector performs better than LSTM and x-vector. For two datasets, the equal error rate(EER) of our Auto-Vector is much lower than LSTM and x-vector. This result proves that the neural architecture search network can find a better model than the expert-designed hand-crafted models. We use the same back-end(GE2E) for all evaluated systems to eliminate the impacts of different back-end classifiers.

The configurations of two baseline networks are shown as follows. The first baseline is a 3-layer \textbf{LSTM Network}\cite{GE2E} with a projection of size 256. The embedding vector(d-vector) size is the same as the LSTM projection size. There are 768 hidden nodes in the LSTM layer. The expected average moving is used to get the embedding. The second baseline is \textbf{x-vector Network}\cite{Xvector}. The first five layers $L_1$ to $L_5$ are frame-level information hidden layers. While there are 512 nodes in each of the first four layers $L_1$ to $L_4$, there are 1500 nodes in the fifth layer $L_5$. The statistical pooling layer builds utterance-level feature by calculating mean and standard deviation over frame-level features. Two utterance-level layers $L_6$ and $L_7$ each have 512 nodes. The sixth layer $L_6$ is used as embedding.

Our \textbf{NAS-based x-vector} is stacked with a repetitive context block. Each block contains 4 choice temporal context windows. As the search space is small, we use random search policy for NAS-based x-vector.

For \textbf{Auto-Vector}, the hyper-parameters of hyper-network (number of blocks $B$ and the number of filters $F$) are analyzed in Section \ref{sec:evaluaitondetails}. The embedding size is set as 512. To decouple the correlation of sub-networks, we set the path dropout rate as 0.1. For the input, the batch size is set as 40 utterances from 8 speakers, each with 5 utterances. For training, we use is Adam optimizer and a linear learning rate decay policy with a base learning rate of 0.02. For the memetic evolutionary search, the size of the population set is 100, and the number of generations is 2000.

\subsection{The Evaluation Details of Auto-Vector}\label{sec:evaluaitondetails}

\begin{table}[]
\caption{The evaluation results of hyper-network and sub-network on Dataset1. Here, $F$ means the number of filters in the first convolution layer and $B$ means the number of choice blocks. For sub-networks, we retrain top-10 sub-networks and report the mean x and standard deviation y as $x\pm y$. }
\label{tab:details}
\center
\begin{tabular}{cccc}
\hline
 &  \begin{tabular}[c]{@{}c@{}}size\\ (M)\end{tabular} &  \begin{tabular}[c]{@{}c@{}}EER\\ (\%)\end{tabular}  & \begin{tabular}[c]{@{}c@{}}cost\\ (GPUh)\end{tabular}\\
\hline
HyperNet(F=16,B=24) & 2.43 & 3.5 & 14.6 \\
HyperNet(F=32,B=24) & 6.08 & 2.7 & 21.7 \\
HyperNet(F=64,B=24) & 17.04 & 1.9 & 33.9 \\
HyperNet(F=128,B=24) &  46.82 & 1.4 & 50.7 \\
\hline
HyperNet(B=12,F=32) & 5.06 & 3.1 &  18.9  \\ 
HyperNet(B=24,F=32) & 6.08 & 2.7 &  21.7  \\ 
HyperNet(B=36,F=32) &  7.08 & 2.6 & 24.3  \\ 
HyperNet(B=48,F=32) & 8.05 & 2.2 & 26.2 \\
\hline
SubNet(F=16,B=48)&$0.8\pm0.2$ & $3.7\pm0.3$ &  -\\
SubNet(F=32,B=48)&$2.1\pm0.5$ & $2.9 \pm0.2$ & -\\
\textbf{SubNet(F=64,B=48)}& \textbf{$6.1\pm1.2$ } & \textbf{$1.9 \pm0.1$} &  - \\
SubNet(F=128,B=48)&$15.2\pm3.2$ & $1.6\pm0.1$ &  -\\
\hline
\end{tabular}
\end{table}

\textbf{Hyper-Network Training}. First, we parameterize our models based on F, the number of filters in the first convolution layer, as shown in Table \ref{tab:details}. When F = 16 and B=24, we obtain an average EER of 3.5 with about 2.43 M parameters. with a double growth of filter number, the growth of model size is multiplied by nearly three times. Obviously, the equal error rate will decrease with the increase of filters. As we use two reduction blocks in our hyper-network model, the growth model size should be multiplied by four times. However, due to the existing of dense layer in the end of NAS-based model, the growth model size is multiplied by nearly three times. The best model gets 1.4\% EER with around 46.82M parameters.  

Then, we parameterize our models based on B, the number of choice blocks. When B= 12 and F=32, we obtain an average EER of 3.1\% with about 5.06M parameters. The best model gets 2.2\% EER with around 8.05M parameters. With a double boost of the block number, the model size only increases a little because there is two dense layers in the tail of our model. The weights of the dense layer dominate the model size, so the model size increases at a low rate along with the double increase of the number of filters. 




\begin{figure}[!htb]
  \centering
    \subfigure[Search hyper-network with F=32 and B=24]{\includegraphics[width=0.4\textwidth]{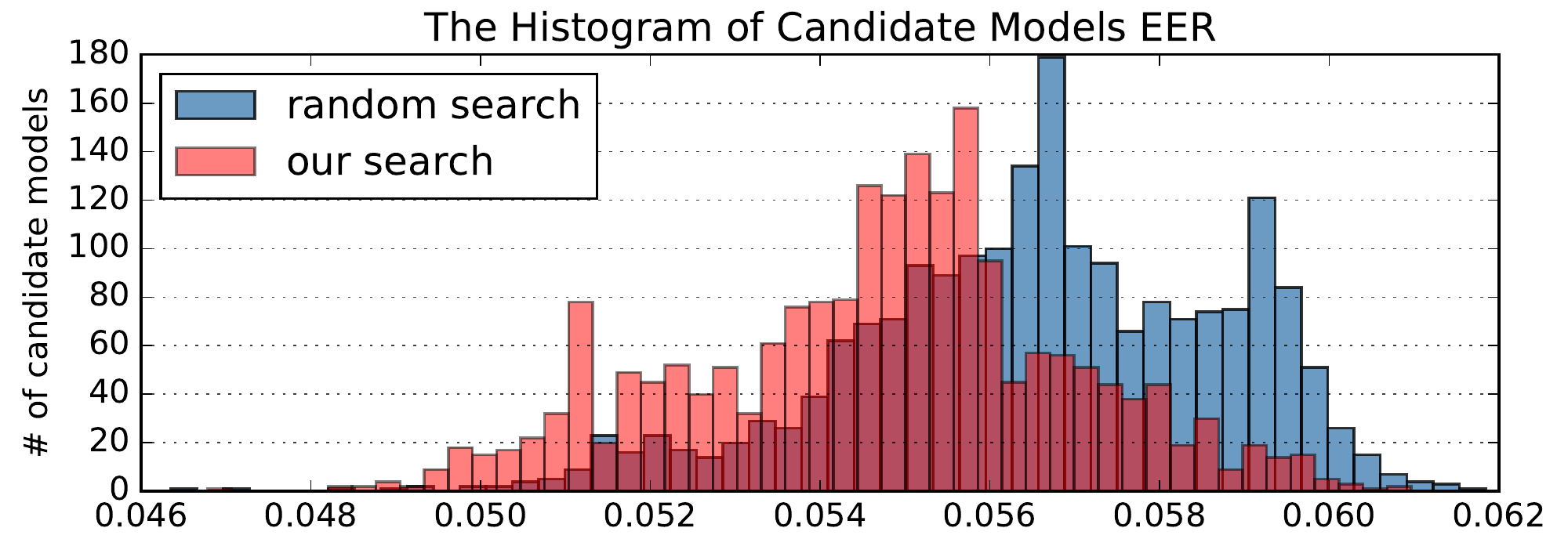}}
    \subfigure[Search hyper-network with F=64 and B=48]{\includegraphics[width=0.4\textwidth]{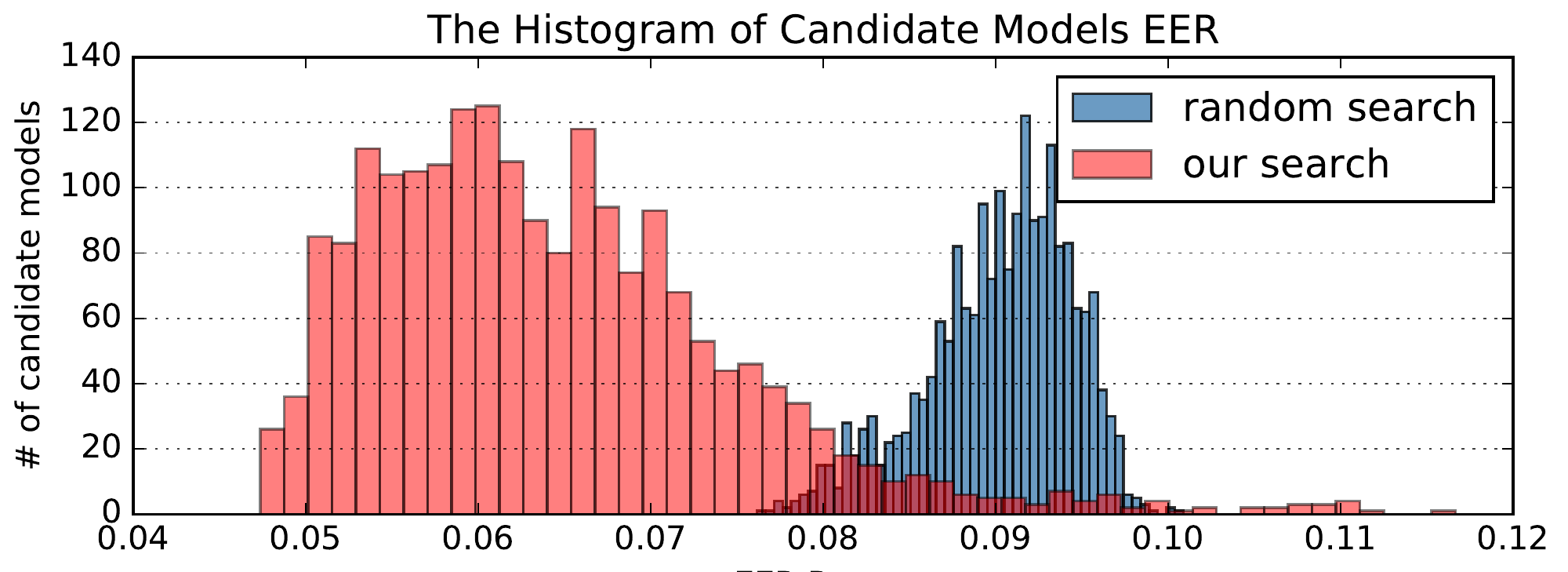}} 
    \caption{The evaluation histogram of 2000 candidates sub-networks with their parameters inherited from the hyper-network} 
    \label{fig:histogram}
\end{figure}

\textbf{The Impact of Evolutionary Search.} Compared to the random selection algorithm, our hierarchical evolutionary algorithm can generate more high-quality candidate models. The equal error rate (EER) distribution of candidate models is shown in Figure \ref{fig:histogram}. While most of the candidate models searched out by our evolutionary algorithm, have a lower equal error rate (EER)  than by random search. Besides, the best-accuracy model is found out by our evolutionary algorithm rather than by a random algorithm. This result further proves that our hierarchical evolutionary algorithm can get more space to be explored to generate more high-quality candidate models. 

\textbf{Sub-Network Re-Training}. As shown in Table \ref{tab:details}, we retrain top-10 sub-networks and report the mean x and standard deviation y as $x\pm y$ for sub-network training. We aim to find the best-quality model whose model size is smaller than the x-vector network. When F=64 and B=48, we can discover the optimal model whose model size is 5.17M, and EER is 1.8\%. 

\section{Conclusion}
In this paper, we introduce the NAS conception into well-known x-vector network. Enabling more search space to be explored, we use an evolutionary algorithm enhanced neural architecture search framework to search high-quality sub-networks. The experiment shows that our system outperforms two state-of-the-art end-to-end methods in a public dataset. Besides, our NAS method can achieve a reduction of 36\%-86\% in equal error compared with the state-of-the-art methods.

\section{Acknowledgment}
This paper is supported by National Key Research and Development Program of China under grant No. 2018YFB1003500, No. 2018YFB0204400 and No. 2017YFB1401202.

\bibliographystyle{IEEEtran}
\bibliography{refs}


\end{document}